\documentclass[aps,pra,reprint,superscriptaddress,letterpaper,twocolumn,longbibliography]{revtex4-2} 
\usepackage[english]{babel}
\usepackage{graphicx}
\usepackage[usenames, dvipsnames, x11names]{xcolor}
\usepackage[colorlinks=true,citecolor=NavyBlue,linkcolor=RubineRed,urlcolor=Cerulean,pdfencoding=auto]{hyperref}
\usepackage{amsmath}
\usepackage{amssymb}

\newcommand{\mm}[1]{\mathrm{#1}}

\def\abs#1{\mathinner{|{#1}|}}

\def\comm#1#2{\mathinner{[{#1},{#2}]}}

\def \hH{\hat{H}}
\def \ha{\hat{a}}

\def \uS{\mathrm{S}}

\hypersetup{
 pdfauthor={Benedikt Tissot},
 pdftitle={Reservoir Engineering for Classical Nonlinear Fields},
 pdfkeywords={},
 pdfsubject={},
 pdflang={English}}

\begin{document}

\begin{abstract}
Reservoir engineering has become a prominent tool to control quantum systems. Recently, there have been first experiments applying
it to  many-body systems, especially with a view to engineer particle-conserving dissipation for quantum simulations using bosons.
In this work, we explore the dissipative dynamics of these systems in the classical limit. We derive a general equation of motion
capturing the effective nonlinear dissipation introduced by the bath and apply it to the special case of a Bose-Hubbard model,
where it leads to an unconventional type of dissipative nonlinear Schrödinger equation. Building on that, we study the dynamics
of one and two solitons in such a dissipative classical field theory.
\end{abstract}

\title{Reservoir Engineering for Classical Nonlinear Fields}
\author{Benedikt Tissot}
\email[]{benedikt.tissot@uni-konstanz.de}
\affiliation{Department of Physics, University of Konstanz, D-78457 Konstanz, Germany}

\author{Hugo Ribeiro}
\email[]{hugo\_ribeiro@uml.edu}
\affiliation{Department of Physics and Applied Physics, University of Massachusetts Lowell, Lowell, MA 01854, USA}

\author{Florian Marquardt}
\email[]{Florian.Marquardt@mpl.mpg.de}
\affiliation{Max Planck Institute for the Science of Light, Staudtstr. 2, 91058
Erlangen}
\affiliation{Department of Physics, Friedrich-Alexander-Universität Erlangen-Nürnberg, Staudstr. 7, 91058 Erlangen}

\maketitle

\section{Introduction}

Reservoir engineering has been originally conceived in quantum physics as a generalization of laser cooling and optical pumping
ideas, representing an approach to harness the coupling to a reservoir which would otherwise just introduce noise and dissipation
\cite{poyatos-1996-quant_reser_engin_with_laser,plenio-2002-entan_light_from_white_noise,
kapit-2017-upsid_noise,harrington-2022-engin_dissip_quant_infor_scien}. By now, the idea of reservoir engineering is already used
in a wide range of platforms, e.g. atoms \cite{krauter-2011-entan_gener_by_dissip_stead}, trapped ions
\cite{barreiro-2011-open_system_quant_simul_with_trapp_ions,lin-2013-dissip_produc_maxim_entan_stead,kienzler-2015-quant_harmon_oscil_state_synth},
optomechanics \cite{wollman-2015-quant_squeez_motion_mechan_reson,pirkkalainen-2015-squeez_quant_noise_motion_microm_reson}, as
well as superconducting circuits
\cite{valenzuela-2006-microw_induc_coolin_super_qubit,murch-2012-cavit_assis_quant_bath_engin,shankar-2013-auton_stabil_entan_between_two,hacohen-gourgy-2015-coolin_auton_feedb_bose_hubbar,ma-2019-dissip_stabil_mott_insul_photon}.
While initially applied to individual quantum systems, recent interest has focussed on many-body scenarios
\cite{wang_PhysRevLett.125.115301,wang2023dissipative,umucal_PhysRevA.104.023704,cian_PhysRevLett.123.063602,ma-2019-dissip_stabil_mott_insul_photon}.

In that context, an important and natural question to ask is whether the dissipation induced by the reservoir can be made to
preserve particle number in the system. In many of the most relevant platforms for quantum simulation \cite{houck2012chip,
altman2021quantum}, this is nontrivial, since excitations of qubits or bosonic particles like photons and phonons are naturally
destroyed by dissipation. At the same time, when particle number is conserved interesting complex ground states can be reached,
e.g., forming a Bose-Einstein condensate or some highly entangled state. Theoretical proposals have investigated that issue in
some detail
\cite{weitz2013optomechanical,hafezi-2015-chemic_poten_light_by_param_coupl,ribeiro-2020-kinet_many_body_reser_engin,harrington-2022-engin_dissip_quant_infor_scien,wang2023dissipative}.

First successful steps have been taken, particularly in an experiment that demonstrated the cooling of a qubit chain to a
Mott-insulating state \cite{ma-2019-dissip_stabil_mott_insul_photon}. As a result, interesting questions have been raised
regarding the kind of many-body states that can be reached using such an approach
\cite{scarlatella2023stability,kurilovich_10.21468/SciPostPhys.13.5.107}. Nonequilibrium Bose condensation of photons is another
important direction making use of particle-conserving dissipation
\cite{klaers-2010-bose_einst_conden_photon_optic_microc,schmitt-2014-obser_grand_canon_number_statis,bloch2022non}.

\begin{figure}
    \centering
    \includegraphics[width=\linewidth]{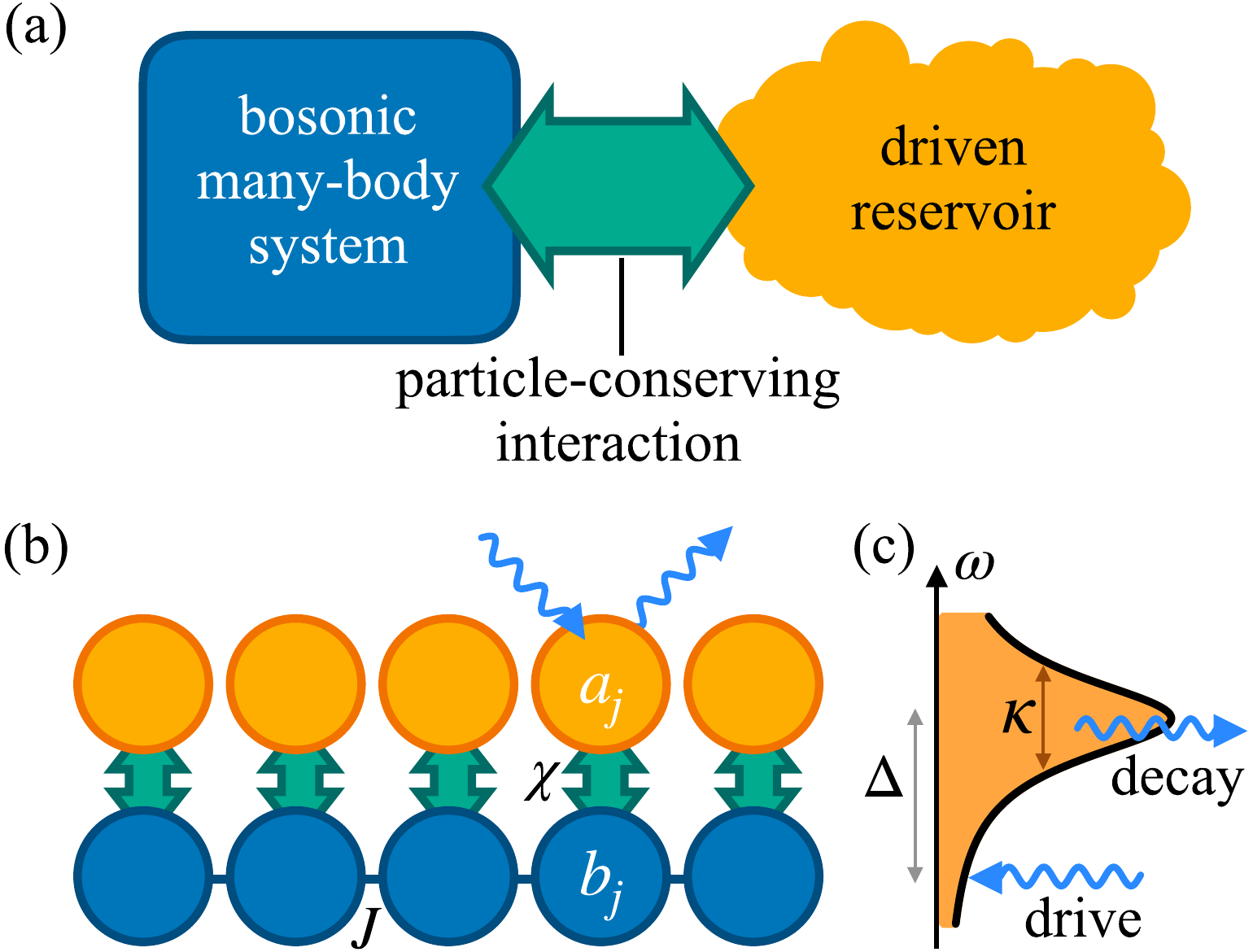}
    \caption{(a) The physical scenario of a bosonic system (in the classical limit) coupled to a driven reservoir, in a
             particle-conserving way. (b) Particular illustrative realization considered in the text, a Bose-Hubbard chain (blue $b_j$) with
             sites coupled to driven cavities (orange bosonic modes $a_j$) with coupling strength $\chi$ and to each other with hopping rate
             $J$. See main text for details. (c) Spectrum of a dissipative cavity, where the incoming drive can be up-scattered in frequency
             through interaction with the bosonic many-body system, extracting energy. The detuning $\Delta$ and the dissipation rate $\kappa$
             determine the properties of this driven reservoir.}
    \label{fig:model}
\end{figure}

When we specialize to a concrete system, we will study the classical limit of a Bose-Hubbard chain where each site is coupled via
its particle number to a driven and dissipative cavity (see Fig.~\ref{fig:model}).  This can potentially be implemented using
superconducting qubits tunnel-coupled to each other, simulating the Bose-Hubbard model
\cite{mansikkamaeki-2021-phases_disor_bose_hubbar_model,fedorov-2021-photon_trans_bose_hubbar_chain}, provided they are
additionally coupled by cross-Kerr interactions to driven microwave resonators forming the reservoir. There are already works
towards particle-conserving interactions with engineered reservoirs, e.g., an experiment showing particle conserving autonomous
cooling of a Bose-Hubbard chain \cite{hacohen-gourgy-2015-coolin_auton_feedb_bose_hubbar} and a recent experiment preparing a Mott
insulator in a Bose-Hubbard chain \cite{ma-2019-dissip_stabil_mott_insul_photon}. Further works in that direction include an
experiment showing Bose-Einstein condensation in an optical cavity \cite{klaers-2010-bose_einst_conden_photon_optic_microc} and
theory work for parametric coupling to generate light with a chemical potential
\cite{hafezi-2015-chemic_poten_light_by_param_coupl}. The particular system we study in the present article has been considered
previously by some of us in the quantum regime, deriving kinetic equations describing the scattering of particles by the coupling
to the reservoir, see Ref.~\cite{ribeiro-2020-kinet_many_body_reser_engin}.

Since many of such systems used for quantum simulations aided by reservoir engineering are bosonic in nature, it is a meaningful
and interesting question to ask how they would behave in the {\em classical limit}. Suppose we have particle-number-conserving
dissipation, brought about by reservoir engineering, and acting on a bosonic many-body system (see Fig.~\ref{fig:model}):~What are
the effective classical equations of motion describing this scenario? Analyzing those may give us new physical insights that can
then even be relevant for interpreting the original quantum dynamics. Moreover, in the classical limit of interacting bosonic
systems, we can find collective excitations like solitons. How do those excitations behave in the presence of particle-number
conserving dissipation? Can we identify how their  motion described in terms of collective variables is modified in the presence
of such unconventional dissipation?

In the present article, we set out to answer these questions. We will present three main results. First, in Sec.~\ref{sec:general}
we derive a general set of classical effective equations of motion for any bosonic many-body system coupled to a driven reservoir
in a particle-conserving manner. These equations are still completely general, in that the Hamiltonian of the bosonic system can
be arbitrary, but they already show the shape of the effective dissipation and allow us to analyze the dependence of the decay
rate on parameters like the detuning between external drive and reservoir modes. The description found there can be viewed as a
generalization of some formulas known from laser-cooling, e.g., in cavity optomechanics, to the case of arbitrary many-body
systems, treated in the classical limit. As a second result, we specialize to a lattice of coupled bosonic modes in
Sec.~\ref{sec:microscopic}, and we derive a nonlinear Schrödinger equation in the presence of particle-conserving dissipation
(both on the lattice and in the continuum limit [Sec.~\ref{sec:PCSE}]). In that equation the dissipation term is nonlinear, in
contrast to the conventional dissipative nonlinear Schrödinger equation, where linear dissipation is considered. Finally, we
apply these equations to describe the motion of solitons. There, as a third main result (Sec.~\ref{sec:colcord}), we show how the
effective equations of motion for the collective coordinates of a soliton are modified by the new dissipation terms. We compare
our analytical derivations with numerical simulations. Lastly, we briefly discuss the interaction of two solitons in
Sec.~\ref{sec:two} and conclude with Sec.~\ref{sec:conclusion}.

\section{System-Reservoir Coupling with Particle Conservation:~Effective Dissipative Equation}
\label{sec:general}

We start out with a general situation, where a bosonic quantum many-body system is coupled to a reservoir in a particle-conserving
manner. This will allow us to derive a formula of wide applicability, describing the effective dynamics of such a system, without
yet specializing to a particular case.
We consider an arbitrary bosonic lattice model. Each system site is coupled independently to a  driven dissipative reservoir mode
${\hat a}_l$. Importantly, to ensure particle-number conservation, the coupling is of the density-density type. Thus, the full
Hamiltonian is of the form
\begin{equation}
    \hH_\mm{tot} = \hH_\mm{S} + \hH_\mm{B} + \hbar \chi \sum_l {\hat a}^{\dagger}_l {\hat a}_l {\hat b}^{\dagger}_l {\hat b}_l.
    \label{eq:Hgeneral}
\end{equation}
Here, $\hH_\mm{S}$ describes the system containing the ${\hat b}_l$ modes, which we do not need to specify further at this point.
The reservoir modes $(\ha_l)$ and their interaction with the environment and external drives
 is contained in $\hH_\mm{B}$. Extensions like correlated reservoirs (couplings of site $l$ to arbitrary
${\hat a}_k^{\dagger} {\hat a}_m$) or disorder in the reservoir modes would be fairly straightforward to implement in the
formalism we are going to discuss and do not change the overall structure.  Couplings between system and reservoir,
with strength $\chi$, of the type postulated in Eq.~\eqref{eq:Hgeneral} arise naturally as cross-Kerr interactions between
optical or microwave modes.

In this article, our goal is to stay in the classical limit of large amplitudes, replacing ${\hat b}\mapsto b$. We will now derive
an effective set of classical equations of motion upon elimination of the reservoir modes.  Let us denote the {\em classical}
Hamiltonian of the {\em system} itself as $H$, which is a function of the classical mode amplitudes $b_l$ and their conjugates
$b_l^*$. The equation of motion in absence of the reservoir is obtained from the normal ordered Hamiltonian $H$ in
general as $i {\dot b}_l = \hbar^{-1} {\partial H / \partial b^*_l}$ (see Appendix~\ref{app:wirtinger} for more details), with the
``Wirtinger derivative'' \cite{wirtinger1927formalen} construction, i.e., assuming $b_l$ and $b_l^*$ to be
independent variables for the purposes of the derivative. The constant $\hbar$ appears in this classical equation only because we
chose to define the classical field amplitudes $b_l$ directly as the limit of the quantum operators (and this retains $\hbar$ as a
conversion factor between energy and frequency). Using that convenient notation, we can now rewrite the dynamical equation in the
form
\begin{equation}
    i \hbar \dot{b}_l = {\partial H_\uS \over \partial b_l^*} + \hbar \chi |a_l|^2 b_l
    \label{eq:classical-full-eq},
\end{equation}
where we denote the temporal derivative using $\dot{b}_l = db_l/dt$. Likewise, there is the equation of motion of the driven
reservoir mode.  Following the standard derivation for a driven, dissipative
mode~\cite{clerk-2010-introd_to_quant_noise_measur_amplif}, we find
\begin{equation}
    i \dot{a}_l = \omega_l a_l + i \eta,
    \label{eq:EOMReservoir}
\end{equation}
with $\omega_l = -\Delta - i \kappa/2 + \chi |b_l|^2$ the effective frequency of mode $a_l$, $\eta$ the strength of the external
drive, and $\kappa$ the energy decay rate. For convenience we have chosen to work in a frame rotating at the
drive frequency  $\omega_L$ of the
reservoir modes and introduced the detuning $\Delta = \omega_L - \omega_{\rm c}$ between the drive and the
frequency  $\omega_\mm{c}$ of a reservoir mode . For simplicity we
assume $\Delta$, $\eta$, and $\kappa$ to be the same for every lattice site $l$.

Due to the dynamics of the system amplitudes $b_l$, the reservoir mode frequency $\omega_l$ will become time-dependent, thus
finding the solutions of Eq.~\eqref{eq:EOMReservoir} is a non-trivial problem. We can, however, look for
perturbative solutions of the form $a_l = \delta a_l - i \eta / \omega_l$, where we chose the second term in analogy to the steady
state of the driven dissipative cavity (for $\chi = 0$) but taking into account the time-dependent frequency shift of the cavity due to
the interaction with the system. To determine  $\delta a_l$, we first
substitute  our ansatz for $a_l$ in the right-hand side of
Eq.~\eqref{eq:EOMReservoir} , which leads to  $i \dot{a}_l = \omega_l \delta a_l$.
Calculating explicitly the time derivative of our ansatz leads to  $i
\dot{a}_l = - \eta \dot{\omega}_l / \omega_l^2 + i \dot{(\delta a_l)}$.  Equating the two obtained expressions
for $i \dot{a}_l$ allows us to find the approximate expression $\delta a_l \approx - \eta
\dot{\omega}_l / \omega_l^3$, where we have neglected the term $\dot{(\delta a_l)}$ which is proportional to $\dot{\omega}_l^2/\omega_l^4, \ddot{\omega}_l/\omega_l^3$.
This is justified within the weak-coupling
limit, where $|\dot{\omega}_l| \ll |\omega_l|^2$ and $\left|(-\Delta - i \kappa/2 - \omega_l)/\omega_l\right| \ll 1$.  In
this form the weak-coupling limit can be ensured via a large decay rate or detuning.

Inserting this approximate solution back into the equation of motion for $b_l$ and expanding up to second order in $\chi$
completes the desired elimination of the reservoir. We note that the expansion in $\chi$ introduces no additional approximations
but simply ensures consistency with the previously employed weak-coupling limit.
The result is the classical effective equation of motion for an {\em arbitrary} bosonic system coupled to a reservoir in a
particle-conserving manner
\begin{equation}
    i \hbar \dot{b}_l = {\partial H_\uS \over \partial b_l^*} + \hbar \delta g |b_l|^2 b_l + \gamma {\mathrm{Im}}\left[ b_l^* {\partial H_\uS \over \partial b_l^*}\right] b_l
    \label{eq:general-main-equation},
\end{equation}
valid in the weak-coupling limit. This is the first main result of the present
work. We   used Eq.~\eqref{eq:classical-full-eq} to get $d |
b_l |^2 /dt =  2 \mathrm{Im} \left[ b_l^* {\partial H / \partial b_l^*}\right]$. The second term on the right hand side of
Eq.~\eqref{eq:general-main-equation} is an effective change of the nonlinearity induced by the coupling to the reservoir with
strength
\begin{equation}
    \delta g = 
    \frac{2 \chi^2 \eta^2 \Delta}{(\kappa^2/4 + \Delta^2)^2}
    \label{eq:effective-nonlin-change}.
\end{equation}
Setting $\Delta<0$ (red detuned drive) in Eq.~\eqref{eq:effective-nonlin-change} leads to an effective attractive interaction. We
have also used an appropriate rotating frame by omitting a trivial static shift of the harmonic frequency   given by  $\hbar
\chi {\eta^2}/{(\Delta^2 + \kappa^2/4)} b_l$ for brevity.

The main effect, which will be the focus of our discussion, is the unusual particle-conserving dissipation represented by the last
term of Eq.~\eqref{eq:general-main-equation}. To verify that the latter corresponds to a dissipative term, we
consider the change in the system's energy given by
\begin{equation}
    \dot{E}_\uS = -\gamma \hbar^{-1} \sum_l \left( \mathrm{Im}\left[ b_l^* {\partial H_\uS \over \partial b_l^*} \right] \right)^2,
    \label{eq:energy-derivative}
\end{equation}
with
\begin{equation}
    \gamma =
    - \frac{4 \chi^2 \eta^2 \Delta \kappa}{(\Delta^2 + \kappa^2/4)^3}.
    \label{eq:gamma-general}
\end{equation}
As long as $\gamma>0$, Eq.~\eqref{eq:gamma-general} predicts that
the energy $E_\uS$
will decrease, confirming the interpretation of the parameter $\gamma$ as a (dimensionless) decay rate.

\begin{figure}[]
    \centering
    \includegraphics[width=\linewidth]{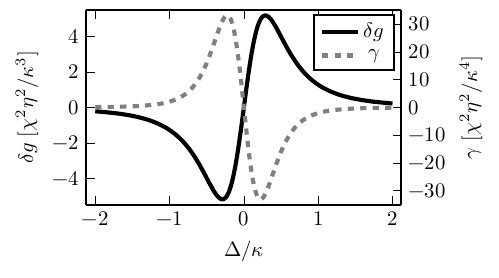}
    \caption{Effective parameters as defined in Eqs.~\eqref{eq:effective-nonlin-change} and \eqref{eq:gamma-general} as functions of the
             detuning:~The correction of the effective onsite interaction \(\delta g\) (black solid) as well as the dissipation parameter
             \(\gamma\) (dashed gray). The dissipation parameter is positive in the red detuned domain (\(\Delta<0\)) and changes sign in the
             blue detuned domain (\(\Delta>0\)). Both the correction of the onsite interaction, as well as the dissipation parameter scale with
             the second order of the bath interaction but have different limiting behaviour at large detuning $\Delta$.}
    \label{fig:effparam}
\end{figure}

Dissipation arises in the red-detuned regime, i.e.,
for $\Delta<0$, which yields a positive $\gamma$. While $\Delta$ allows one to control the sign of
$\gamma$, the strength of $\gamma$ depends on the interplay between the drive strength $\eta$, the decay rate
$\kappa$ of the reservoir mode, and the coupling $\chi$ to the system. We show in Fig.~\ref{fig:effparam} the dependence on
the effective parameters.

Even though we concentrate on classical dynamics, it is insightful to understand the origin of the decay
mechanism  in the original quantum mechanical picture, where the incoming red-detuned photon is up-scattered to the higher
resonance frequency of the reservoir mode by absorbing an energy excitation of the bosonic system, thereby leading to dissipation.
At the same time, the total particle number, $N=\sum_l |b_l|^2$ is conserved if \(H_{S}\) conserves \(N\), which can be readily verified  by taking the time derivative of $N$ and employing Eq.~\eqref{eq:general-main-equation}.

Formulas like Eq.~\eqref{eq:gamma-general} are known from the theory of laser cooling \cite{wineland1975proposed}, and its recent
implementations , e.g., in cavity optomechanics \cite{aspelmeyer2014cavity}. Indeed there is a
direct connection of the particle-conserving coupling studied here to quadratic optomechanics, where one couples the light mode to
the square of the mechanical displacement \cite{thompson2008strong,jayich2008dispersive,nunnenkamp2010cooling}, albeit now with
the mechanical oscillator replaced by a many-body system.

\section{Microscopic Model:~The Anharmonic Chain}
\label{sec:microscopic}

To derive the dynamical consequences of Eq.~\eqref{eq:general-main-equation}, we will in the following invoke a specific
illustrative example, namely a chain of anharmonic bosonic modes, corresponding to the classical limit of the Bose-Hubbard chain.
This will already produce very rich behaviour which we are able to analyze both numerically and semi-analytically.

The classical system Hamiltonian for this chain is
\begin{equation}
    H_{\mathrm{chain}}/\hbar = -\sum_{l} J (b^{*}_{l+1} b_l + {\rm c.c.}) + \frac{\alpha}{2} b_l^{*}b_l (b_l^*b_l -1)
    \label{eq:anharmonic-chain},
\end{equation}
where $J$ is the hopping amplitude between the sites of the chain  and $\alpha$ denotes the
anharmonicity (negative for attractive interactions).  Such a chain can describe various systems:~Bosonic cold atoms in a 1D
lattice with a Hubbard interaction, coupled superconducting transmon qubits (with $\alpha<0$ in that case), as well as coupled
optical cavities with a Kerr interaction inside the cavities.  As we will explore in some detail below, its continuum limit can
give rise to solitonic solutions. Without loss of generality we assume $J$ to be positive in
Eq.~\eqref{eq:anharmonic-chain}~\footnote{ By changing from $b_l \to \tilde{b}_l = (-1)^l b_l$ the sign in front of the hopping
term flips, such that we can choose a description where $J$ is positive.}

\subsection{Nonlinear Schrödinger Equation with Particle-Conserving Dissipation}
\label{sec:PCSE}

We now focus on the 1D anharmonic chain described by Eq.~\eqref{eq:anharmonic-chain}.
To obtain an equation describing the effective dynamics of the system, we use Eq.~\eqref{eq:general-main-equation} with $H_\uS = H_{\mathrm{chain}}$.
This  yields an effective nonlinear Schrödinger equation on a lattice, which includes particle-conserving dissipation (in contrast to the nonlinear Schrödinger
equation with linear loss~\cite{rossi-2020-non_conser_variat_approx_nonlin_schroed_equat}):
\begin{equation}
    \begin{aligned}
        i \dot{b}_n = & g |b_n|^2 b_n - J \left( b_{n-1} - 2 b_n + b_{n+1} \right) \\
        &  - J \gamma \mathrm{Im} \left[ b_n^{*} \left( b_{n-1} - 2 b_n + b_{n+1} \right) \right] b_n ,
    \end{aligned}
\label{eq:dbEFF}
\end{equation}
where $\gamma$ is given by Eq.~\eqref{eq:gamma-general}, $g=\alpha + \delta g$ is the renormalized coupling strength
of the nonlinearity due to the reservoir [see Eq.~\eqref{eq:effective-nonlin-change}], and $n=1, \ldots, L$ labels the lattice
sites. In the following, we consider periodic boundary conditions, i.e., $b_j = b_{j+L}$ (\(j \in \mathbb{Z}\)). In our analytic
derivation we further consider open boundary conditions where $b_0 = b_{L+1} = 0$. An alternative derivation of
Eq.~\eqref{eq:dbEFF} is presented in Appendix~\ref{app:cavity} where we start from the general exact solution.

\begin{figure}[]
    \centering
    \includegraphics[width=\linewidth]{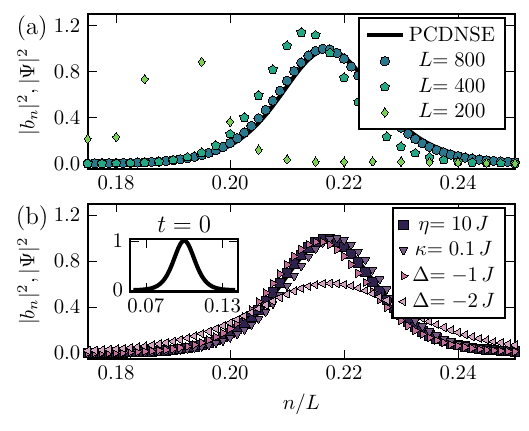}
    \caption{
             Comparison of the PCDNSE [Eq.~\eqref{eq:nls}] and the discrete equations of motion including the reservoir dynamics, see Eqs.~\eqref{eq:classical-full-eq} and \eqref{eq:EOMReservoir}.
             In panel (a) we compare simulations for different array lengths to the PCDNSE and in (b) different cavity parameters leading to the same effective evolution.
             Only when the continuum limit or weak-coupling limit are not satisfied (small $L$ or \(\Delta = -2 \, J\), respectively) the discrete dynamics are in disagreement with the dynamics of the PCDNSE.
             Unless otherwise specified  (see legend) we use \(L=800, \kappa = J, \Delta=-0.1 J, \eta= J\) and
             $\chi,\alpha$ such that we get the effective parameters \(g=-0.1\,J, \gamma=0.05\).
             The distributions are taken after an evolution for \(J t = 50 (L/400)^2\) of an initial stable soliton [see inset of (b)] with height \(\psi(0) = 1\) and velocity \(v(0) \approx 0.48\). The PCDNSE uses a length of \(L=400\).
             For the evolution according to the Langevin EOM, we assume that the cavities are initially in their unperturbed steady state $\eta / (\kappa/2 - i \Delta)$.
        }
    \label{fig:cont}
\end{figure}

Within the effective dynamics, the amount of relevant parameters is reduced from formerly six to three, with $J$ setting the
overall frequency scale and the dimensionless parameters (\(g/J,\,\gamma\)) determining the qualitative behaviour (see also
Fig.~\ref{fig:cont} for an illustration on how different microscopic parameters can lead to the same effective dynamics).

For large arrays (\(L \gg 1\)) and assuming that the lattice spacing \(\delta x=1\) is smaller than typical
wavelengths, i.e., the solution varies slowly on the scale of the lattice spacing, it is possible to consider the continuum limit
of Eq.~\eqref{eq:dbEFF}. To this end, we introduce the continuous function \(\Psi(x, t)\) with \(x \in [0,L]\), which obeys
\(\Psi(x=n,t) = b_n(t)\) for \(n=1,\dots,L\). To obtain the continuous limit of Eq.~\eqref{eq:dbEFF}, we replace the finite
difference \({[\Psi(x+\delta x) - 2 \Psi(x) + \Psi(x-\delta x)]}/{\delta x^2}\) by the second order derivative of $\Psi(x)$ with
respect to the position $x$ \({\partial^2 \Psi}/{\partial x^2}\). This yields the \emph{particle-conserving dissipative nonlinear
Schrödinger equation} (PCDNSE):
\begin{equation}
    i \dot{\Psi} = g |\Psi|^2 \Psi - J \frac{\partial^2 \Psi}{\partial x^2} - J \gamma \mathrm{Im} \left( \Psi^{*} \frac{\partial^2 \Psi}{\partial x^2} \right) \Psi ,
    \label{eq:nls}
\end{equation}
with the parameters $\gamma$ [see Eq.~\eqref{eq:gamma-general}], $g$, and $J$ defined above [see Eqs.~\eqref{eq:gamma-general} and
\eqref{eq:dbEFF}].
Particle conservation here implies that  \(\int dx \abs{\Psi}^2(x,t) = N\) is constant (see Appendix~\ref{app:cons}).
Equation~\eqref{eq:nls} constitutes the second main result of our work.
It will form the basis for our analysis in the remainder of this work, deriving the physical consequences of particle-conserving dissipation for this
paradigmatic 1D model.

Since the nonlinear Schrödinger equation without dissipation is known to harbour solitons, we investigate their dynamics in the
presence of particle-conserving dissipation. For this purpose, the sign of the
dissipation parameter \(\gamma\), which is positive (negative) for red (blue) detuning, is of the utmost importance [see
Eq.~\eqref{eq:gamma-general}].  We show in the following that in the red-detuned regime, wave packets in the form of soliton
solutions can be stabilized and de-accelerated. In contrast to the red-detuned regime, they are
accelerated in the blue-detuned regime and any perturbations away from stable solutions can be amplified,
leading to wave packets that can break up into several parts. In the
following  we mainly focus on the dynamics in the red-detuned regime.

To verify the validity  of Eq.~\eqref{eq:nls} we simulate both the original equations
describing the classical dynamics of a discrete lattice of bosonic modes coupled to driven cavities [see Eqs.~\eqref{eq:classical-full-eq},~\eqref{eq:EOMReservoir} and for the Hamiltonian~\eqref{eq:anharmonic-chain} as well as
Appendix \ref{app:cavity}] and the
continuum particle-conserving dissipative nonlinear Schrödinger equation given by  Eq.~\eqref{eq:nls}.
We compare the resulting dynamics  for several different parameter choices.

As an initial condition we choose a stable soliton, that will be introduced in more detail in Sec.~\ref{sec:colcord}. We set up
the simulation  such that the extent of the soliton is some fraction of the total chain length $L$,
and we keep this fraction constant while we vary $L$ in order to discuss the deviations from the continuum limit (which is
attained perfectly when $L\rightarrow \infty$).

In Fig.~\ref{fig:cont} we compare the dynamics of the discrete model before eliminating the reservoir modes
 [see Eqs.~\eqref{eq:classical-full-eq},~\eqref{eq:EOMReservoir}~and~\eqref{eq:anharmonic-chain} as well as Eqs.~\eqref{eq:dA}~and~\eqref{eq:dB} for
further details] with the PCDNSE.
We plot the solutions for
different array lengths $L$ [see Fig.~\ref{fig:cont}(a)] and different reservoir parameters [see Fig.~\ref{fig:cont}(b)].
Our results show  that for long arrays, $L
\gg 1$, and within the weak-coupling limit both dynamics are in good agreement. Due to the intensity-intensity coupling to the reservoir the site occupation needs to be accounted for in the effective coupling strength such that the weak-coupling limit corresponds to having
\(\chi {b_{\mathrm{max}}}^2 \ll \abs{i \Delta + \kappa/2}\) and \(\chi J {b_{\mathrm{max}}}^2 \ll \abs{i \Delta + \kappa/2}^2\),
where we have defined $b_{\mathrm{max}} = \max_l |b_l|$. We note that $b_{\mathrm{max}}$ can be related to the
occupation of the soliton divided by the width of the soliton. As it can be observed in
Fig.~\ref{fig:cont}(b) if the weak coupling conditions are not fulfilled, the discrete
and continuous solutions do not agree anymore because integrating out the cavities is no longer justified.

As anticipated, the continuum approximation becomes worse for solitons that
(initially) extend only over a few sites, e.g., about $10$ sites for  $L=400$.
However, even in this case  the continuum equation already agrees reasonably well
with the site occupations of its discrete counterpart [see Fig.~\ref{fig:cont}(a)].
We stress  that within the continuum and weak-coupling limits different parameters that produce
the same effective nonlinear coupling strength $g$ and decay rate $\gamma$ lead to equivalent dynamics.
For this reason, we focus in the following on a better
understanding of the dynamics of the continuum equation [see Eq.~\eqref{eq:nls}].

\subsection{Variational Ansatz}
\label{sec:colcord}

Employing the PCDNSE allows us to use  {(semi-)analytic} methods. In particular we use a
variational approach  to describe the dynamics of collective coordinates
\cite{scott-2006-nonlinear,dawes-2013-variat_approx_use_collec_coord,sahoo-2017-pertur_dissip_solit,rossi-2020-non_conser_variat_approx_nonlin_schroed_equat},
here applied to obtain the dynamics of a single soliton. We consider the ansatz
\begin{align}
        \bar{\Psi}(x,t) = \exp\Big(i\Big\{ & \left[x - {x_{0}}{(t)}\right] v{(t)} + \left[x - {x_{0}}{(t)}\right]^{2} d{(t)}  \notag \\
& + \varphi{(t)}\Big\}\Big)
                          \psi{(t)} \operatorname{sech}{\left[\frac{x - {x_{0}}{(t)}}{w(t)} \right]},
    \label{eq:ans}
\end{align}
with the collective coordinates given by the amplitude \(\psi(t)\), the center position \(x_0(t)\), the velocity \(v(t)\), the
width \(w(t)\), as well as a parameter quantifying the (quadratic) phase dispersion \(d(t)\) and the global phase \(\varphi(t)\).
A straightforward calculation (see Appendix~\ref{app:var}) yields the equations of motion for the collective coordinates (EOMs):
\begin{align}
  \label{eq:EOMs1}
  \dot{x_{0}}{(t)} =\, & 2 J v{(t)} , \\
  \label{eq:EOMs2}
  \dot{v}{(t)} =\, & - \frac{8 J \gamma \psi^{2}{(t)} v{(t)}}{15 w^{2}{(t)}} , \\
  \label{eq:EOMs3}
  \dot{\psi}{(t)} =\, & - 2 J \psi{(t)} d{(t)} , \\
  \label{eq:EOMs4}
  \dot{w}{(t)} =\, & 4 J d{(t)}w{(t)} , \\
  \label{eq:dd}
  \dot{d}{(t)} =\, & - \frac{8 J \gamma \psi^{2}{(t)} d{(t)}}{15 w^{2}{(t)}} - 4 J d^{2}{(t)} + \frac{4 J}{\pi^{2} w^{4}{(t)}} \notag \\
  &+ \frac{2 g \psi^{2}{(t)}}{\pi^{2} w^{2}{(t)}} , \\
  \dot{\varphi}{(t)} =\, & \frac{2 \pi^{2} J \gamma \psi^{2}{(t)} d{(t)}}{45} + \frac{2 J \gamma \psi^{2}{(t)} d{(t)}}{3} + J v^{2}{(t)} \notag \\
  \label{eq:EOMs6}
 & - \frac{2 J}{3 w^{2}{(t)}} - \frac{5 g \psi^{2}{(t)}}{6} .
\end{align}
The effects due to reservoir engineering emerge from
all the terms proportional to \(\gamma\). When setting $\gamma=0$, we can verify that these equations coincide with those found in
the literature for soliton dynamics \cite{rossi-2020-non_conser_variat_approx_nonlin_schroed_equat}.

As a first step in our analysis, we determine the
non-trivial stable soliton using Eqs.~\eqref{eq:EOMs1}-\eqref{eq:EOMs6}. To find the properties of the stable soliton, we impose
both the width and the amplitude to remain stationary:~\( \dot{w}{(t)} = \dot{\psi}{(t)} = 0\). This condition implies that the
dispersion vanishes, i.e., \(d(0) = \dot{d}{(t)} = \frac{4 J}{\pi^{2} w^{4}{(t)}} + \frac{2 g \psi^{2}{(t)}}{\pi^{2}
w^{2}{(t)}}=0\). This immediately leads to a  relation between width and amplitude of the stable soliton given by
\begin{equation}
    \psi^2{(t)} = - \frac{2 J}{g w^2{(t)}},
    \label{eq:psiS}
\end{equation}
and imposes the quadratic phase dispersion to be zero, \(d{(t)} = 0\).
The initial conditions of the remaining parameters \(v{(t=0)},x_0{(t=0)},\varphi{(t=0)}\) can be chosen independently.

It can be shown that both the width and amplitude of the stable soliton are time-independent. To this end, we use that the particle number,
\begin{equation}
    N = \int dx \abs{\Psi}^2(x,t) = 2 \psi^2 w,
    \label{eq:Nvar}
\end{equation}
is conserved.
Substituting Eq.~\eqref{eq:psiS} into Eq.~\eqref{eq:Nvar} and solving for $w$ leads to the width of the stable soliton:
\begin{equation}
        w_{\mathrm{SS}} = - \frac{4 J}{g N}.
    \label{eq:wSS}
\end{equation}
The time-independent amplitude can be obtained by substituting Eq.~\eqref{eq:wSS} into Eq.~\eqref{eq:Nvar} and solving for $\psi$, which yields
\begin{equation}
        \psi_{\mathrm{SS}} = \sqrt{\frac{N}{2 w_{\mathrm{SS}}}} = \sqrt{- \frac{g}{2 J}} \frac{N}{2}.
    \label{eq:psiSS}
\end{equation}

We can also derive the shape of the stable soliton using the ansatz postulated in
Eq.~\eqref{eq:ans}; this would consist in  minimizing the energy of the unperturbed
nonlinear Schrödinger equation for a fixed particle number
\begin{equation}
    \begin{aligned}
        E = & \int dx \left[ J \abs{\frac{\partial \Psi}{\partial x}}^2 + \frac{g}{2} \abs{\Psi}^4 \right] \\
          = & 2 J \psi^{2} v^{2} w + \frac{2 \pi^{2} J \psi^{2} d^{2} w^{3}}{3} + \frac{2 J \psi^{2}}{3 w} + \frac{2 \psi^{4} g w}{3}.
    \end{aligned}
    \label{eq:E_NSE}
\end{equation}
In the
attractive regime, i.e., $g<0$, the last term of Eq.~\eqref{eq:E_NSE} related to the onsite interaction is negative and, thus,
competes with the remaining positive terms allowing for a stable soliton. Furthermore, Eq.~\eqref{eq:E_NSE} reveals the dependence
of the energy of the stable soliton on the collective variables, which demonstrates that it is possible to change the energy of the
stable soliton without breaking it apart by changing its shape (amplitude $\psi$ and width $w$) or reducing the
dispersion \(d\) and or velocity \(v\).
As we show below, this can be achieved by leveraging the coupling to the engineered
reservoir.
By substituting $w = N / 2\psi^2$ [Eq.~\eqref{eq:Nvar}] into the energy Eq.~\eqref{eq:E_NSE} we find that the kinetic energy term of the group velocity $v$ is independent of the soliton shape (determined by $\psi$).
Minimizing Eq.~\eqref{eq:E_NSE} with respect to $\psi$ and $d$ for attractive onsite interaction (\(g < 0\)) at fixed $N$ and $v$ leads to
\(d=0\), and Eqs.~\eqref{eq:wSS} and \eqref{eq:psiSS}.
The particular shape of the soliton [Eqs.~\eqref{eq:wSS} and \eqref{eq:psiSS}] arises due to the competition between part of the hopping term and the onsite interaction,
in particular the third and last terms of Eq.~\eqref{eq:E_NSE}, respectively.

The energy of the stable soliton is given by
\begin{equation}
    E_{\mathrm{SS}} = J v^{2} N - \frac{N^3 g^2}{48 J}.
    \label{eq:Establesoliton}
\end{equation}
The first term of Eq.~\eqref{eq:Establesoliton} is the kinetic energy which depends quadratically on the  velocity
\(v\). The second term corresponds to a potential energy that depends on the particle number $N$ as well as the system
parameters $g$ and $J$. The energy of the stable soliton [see Eq.~\eqref{eq:Establesoliton}] is linked to the evolution of the
global phase via $dE_{SS}/dN = v \dot{x_0}- \dot{\varphi}$ [see Eqs. \eqref{eq:EOMsStable1} and \eqref{eq:EOMsStable3}].

\subsubsection{Dissipative dynamics of the stable soliton}
\label{sec:dissdynsoliton}

With these considerations in place we begin by studying the dynamics of a stable soliton. Solitons in Bose-Hubbard-type models
have been discussed in the quantum domain \cite{blain2023soliton,naldesi_PhysRevLett.122.053001} without the particle-conserving
dissipation we consider  here. Because the PCDNSE conserves the particle number but dissipates the energy, the dynamics need to
change the velocity to influence the energy as it can be seen from Eq.~\eqref{eq:Establesoliton}.  This can be
simply understood using the EOMs [see Eqs.~\eqref{eq:EOMs1}-\eqref{eq:EOMs6}] for a stable soliton
\begin{align}
  \label{eq:EOMsStable1}
  \dot{x_{0}}{(t)} & = 2 J v{(t)} , \\
  \label{eq:EOMsStable2}
  \dot{v}{(t)} & = - \Gamma J v{(t)}, \\
  \label{eq:EOMsStable3}
  \dot{\varphi}{(t)} & = J v^{2}{(t)} + \frac{g^2 N^2}{16 J} ,
\end{align}
with the (dimensionless) velocity damping rate given by
\begin{equation}
  \Gamma = \gamma \frac{- g^3}{J^3} \frac{N^4}{240}.
  \label{eq:velocityDamping}
\end{equation}
Note that the velocity damping rate of the stable soliton depends on the ratio \(g/J\) because this fraction determines the
particular shape of the stable soliton, i.e., the amplitude $\psi$ and width $w$ [see Eq.~\eqref{eq:EOMs2}].

Since we need an attractive interaction ($g<0$) to get a stable soliton, we conclude that the velocity damping rate $\Gamma$ is
positive when the original effective damping parameter $\gamma$ is also positive, i.e., when $\Delta < 0$ (red-detuned regime).
Therefore, the main insight these equations provide is that the velocity \(v\) of the soliton is damped at a rate of \(\Gamma J\)
for $\Delta<0$. This highlights that with the particle-conserving coupling to the engineered reservoir it is indeed possible to
de-accelerate and eventually even freeze the motion of a stable soliton while conserving its shape. On the other
hand when  $\Delta>0$, one can observe instead exponential acceleration.

\begin{figure}
    \centering
    \includegraphics[width=\linewidth]{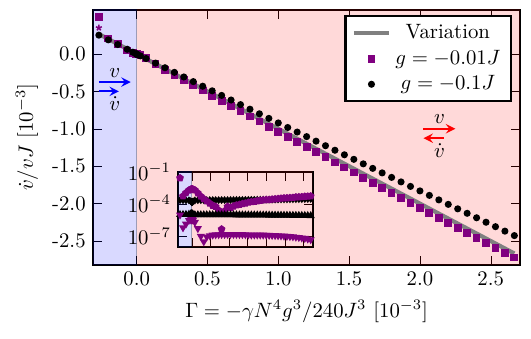}
    \caption{Velocity damping rate as a function of \(g \gamma \psi^4(0)\). We see good agreement between the variational approach
        and numerical results of the particle conserving dissipative nonlinear Schrödinger equation (PCDNSE). The deviation at the
        smallest value is due to the soliton breaking up in the simulation, underlined by the datapoint given by the star where
        the relative and absolut tolerances where decreased to \(10^{-13}\) and \(10^{-12}\) respectively from \(10^{-8}\) for
        both.The inset shows that the deviation \(|\delta(Jt=4)| = | \psi^{2}(Jt=4)/\psi^{2}(0) - 1 |\) (pentagons, diamonds) as well
        as \(d(Jt=4)\) (triangles) from the stable soliton stays small for the evolution.  The PCDNSE simulation is for a stable soliton
        with \(\psi(0) = 1\), $x(0) = L/8$, $v(0) \approx 0.49$, and $L=600$ and for a duration $Jt = 4$.  }
    \label{fig:velocity}
\end{figure}

We show this behavior  in Fig.~\ref{fig:velocity} where we compare the relative
(de-)acceleration of the soliton $\dot{v}/vJ$ as function of the damping rate $\Gamma$ as predicted by the
variational ansatz to a numeric simulation of the PCDNSE.  We simulate
numerically the evolution up to $Jt=4$ and then extract the soliton coordinates using a least squares fit to the
ansatz given in Eq.~\eqref{eq:ans}. We then estimate $\dot{v}/J$ using the  finite difference
$[v(Jt) - v(0)]/Jt$.  We stress that while in the fully displayed red-detuned domain the numerical results agree nicely with the
predicted linear dependence on the damping rate $\Gamma$, in the blue-detuned regime there is a sudden increase of the deviation
from  the predicted linear dependence (leftmost purple square in Fig.~\ref{fig:velocity}). We attribute this
deviation to the amplification of a small departure  from the
stable soliton, which is possible in the blue-detuned domain [see Eq.~\eqref{eq:dd}].

In the simulation numerical errors can lead to a small deviation from the stable soliton which can result in a non-vanishing dispersion $d \ne 0$ that then gets amplified for $\gamma < 0$ [see Eq.~\eqref{eq:dd}].
We verified this by reducing the absolute and relative tolerances of the simulation and thereby the potential deviation from the stable soliton, resulting in  final numerical values for the
anti-damping rate in agreement with the variational calculation (see purple star in Fig.~\ref{fig:velocity}).
Due to the amplification, even small deviations from the stable soliton can eventually lead to a break-up of the soliton.  As (external) perturbation can also occur in the real world, the
blue-detuned regime needs to be treated with great care if the goal is to prepare stable solitons.

\subsubsection{Shape stabilization}

\begin{figure*}
    \centering
    \includegraphics[width=\linewidth]{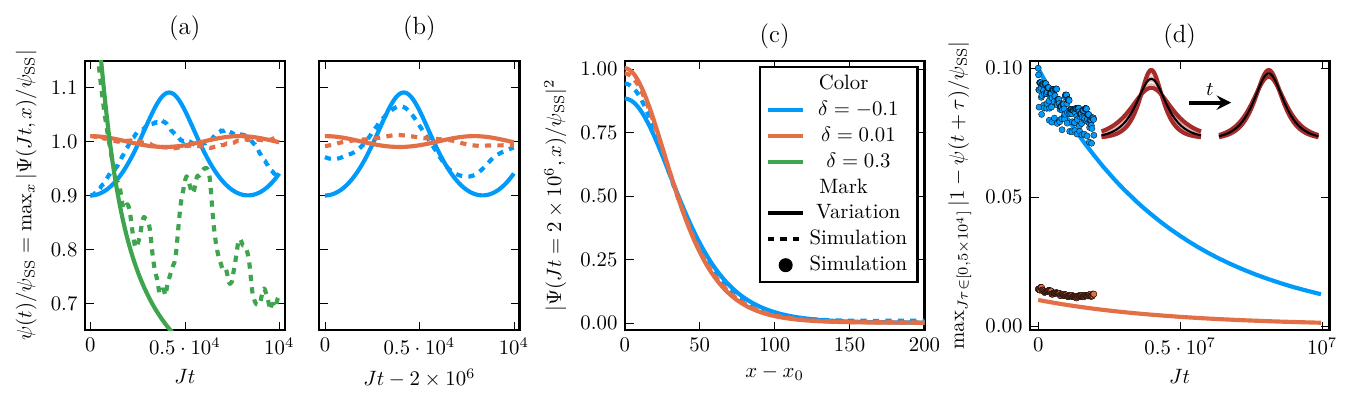}
    \caption{Shape stabilization of \emph{ill} prepared solitons.  The maximal amplitude of the PCDNSE (dashed lines) is compared
        to the prediction of the collective coordinates (solid lines) in (a),(b) for the different deviations $\delta = \psi(0) /
        \psi_{\mathrm{SS}} - 1$ [see legend in (c)].  While for $\delta=-0.1,0.01$ both solutions oscillate around
        the steady state amplitude, for $\delta=0.3$ the soliton shape breaks down and this oscillation is visible in neither
        solution.  Therefore, we investigate the dynamics for long times for the smaller $\delta=-0.1,0.01$ (b)--(d).  Panel (b)
        shows the the continued oscillation around the steady state amplitude after a prolonged evolution
        time of $Jt = 2 \times 10^6$ and (c) displays the modulo square distribution after this evolution.  To quantify the long time dynamics we
        compare the envelope of the deviation of the maximal amplitude $\psi(t)$ deviation from the steady state amplitude
        $\psi_{\mathrm{SS}}$ on a long timescale in these cases (d).  To this end, we calculate the envelope as the maximal
        deviation within a time-window of length $5\times10^4$.   The relaxation process is sketched in the inset of (d).
        We use \(g = -0.1 J\), \(N=1\), \(\gamma=0.1\), \(L=10w_{\mathrm{SS}}\), and only deviate from the stable soliton via
        \(\delta \ne 0\) (i.e. \(d=0\)).  }
    \label{fig:shape}
\end{figure*}

In contrast to the shape breakdown in the blue-detuned regime, which we discussed   above, the
red-detuned domain supports the stabilization of solitons when they start off with a small deviation from the stable shape. We now
analyze this behaviour in some more detail. Since the EOMs conserve the particle number, the initial deviation from a stable soliton can be quantified by the initial phase dispersion \(d(0)\) and initial
relative difference in amplitude \(\delta = \psi(0) / \psi_{\mathrm{SS}} - 1\).  For simplicity we focus on the second type of
deviations here, i.e., deviations in the initial amplitude.

To study the phenomenon of soliton stabilization, we need to consider the long-time limit of the evolution. While
this can be done efficiently using the variational approach, we must make sure that we are not in a regime where small
amplitude deviations from the stable soliton characterized by $\delta$ result in the wave packet breaking up into multiple components. To identify the
regime of validity, we employ a hybrid approach:~We first calculate a short-time evolution using the computationally more
expensive PCDNSE to ensure the stability of the soliton. If the soliton retains the correct shape, we then use the collective
coordinates to estimate the long time evolution.

We illustrate and justify this approach in Fig.~\ref{fig:shape} where we compare the PCDNSE and variational
approach. Based on the short-time evolution of $\psi(t)$ in Fig.~\ref{fig:shape}(a),
one can identify  that the soliton breaks up into multiple wave
packets for $\delta=0.3$ within the PCDNSE. Thus, we conclude that the variational approach is not suited
 in this case.

For the remaining values of $\delta =-0.1,0.01$
we additionally compare the dynamics on a short time scale (after a previous long evolution)
in panel (b) and verify that the PCDNSE solitons still agree with the spatial form of the general soliton ansatz \eqref{eq:ans} in
panel (c).
Both underline the validity of the hybrid approach.

In Fig.~\ref{fig:shape}(d) we plot the maximal deviation of the soliton amplitude from the steady state value within a timespan of
duration $5 \times 10^4/J$ to confirm that for red-detuned dynamics the solitons converge to a state
where the velocity \(v\) and dispersion \(d\) tend to zero, while the amplitude $\psi(t)$ and width \(w(t)\) tend to
the values associated to a stable soliton, i.e., $\psi_{\mathrm{SS}}$ and $w_{\mathrm{SS}}$ [see Eqs.~\eqref{eq:wSS} and \eqref{eq:psiSS}].

Even for the  smaller $|\delta|$ values, the PCDNSE simulation  and the
collective coordinates approach also deviate due to higher order dispersion terms that are not accounted for in our
collective coordinate ansatz.  Because the higher order dispersion terms do not influence the (modulus square) shape they can be
incorporated in the variational approach to improve the collective coordinates to higher order in future work.  The simulations in
Fig.~\ref{fig:shape} furthermore suggest that the ansatz we employed still gives an estimate for the timescale of the relaxation
as well as the qualitative dynamics of the relaxation.

The relevance of the hybrid approach becomes evident in the blue-detuned regime even in the case where the initial
soliton is stable; in this regime a slight perturbation from the stable soliton solution can be amplified and lead to a break up
of the wavefunction as we discussed in the previous section.
Additionally, the blue-detuned regime reveals  the main drawback of the variational approach employed
here, i.e., the complete breakdown of the solution as soon as the space spanned by the ansatz is insufficient to describe the
dynamics. However,  we want to stress again the great simplification that is possible as
long as the solutions stays within this solution space,
i.e.~replacing the non-linear partial differential equation~\eqref{eq:nls} with a set of ordinary differential equations~\eqref{eq:EOMs1}--\eqref{eq:EOMs6}.
This enabled our analytic results
like the derivation of the velocity damping rate for a stable soliton [see
Eqs.~\eqref{eq:EOMsStable1}--\eqref{eq:EOMsStable3} and Fig.~\ref{fig:velocity}].

We emphasize  that the breakdown of the collective-coordinates approach is a problem inherent to taking
the highly restricted ansatz given by Eq.~\eqref{eq:ansAPP}, and is not due to the particular application to the
model of this article; this is particularly visible as the approach works better for bigger \(\gamma\) (for fixed \(g/J\)).

Combining the insights gained above for the red-detuned domain, regarding the velocity damping and shape stabilization, as well as
the separate finding that large deviations lead to a break up of the wavepacket, we suspect that the eventual
steady states for the break up scenario can be described by several spatially separated stable solitons.

\subsection{Two Solitons}
\label{sec:two}

After investigating single solitons in great detail in the previous section we want to use these insights now to
investigate the collision of two solitons.
To this end, we investigate the effect of the interaction by comparing the evolution of a single traveling soliton
using the solution obtained using the collective coordinates to the evolution of two solitons propagating in opposite directions,
evolved (numerically) using the PCDNSE [using Eq.~\eqref{eq:E_NSE}]. In particular, we calculate the
total energy numerically and compare it to the energy of a single soliton, calculated analytically according to the collective
coordinates ansatz [see Eqs.~\eqref{eq:Establesoliton}~and~\eqref{eq:EOMsStable2}].

\begin{figure}[t]
    \centering
    \includegraphics[width=\linewidth]{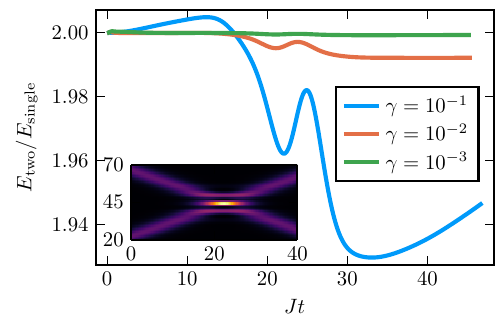}
    \caption{Influence of the dissipative dynamics due to the interaction of two stable solitons. The figure depicts the energy of
        the two solitons compared to the single soliton solution as a function of time during the interaction for different
        \(\gamma\) (see legend).  The inset shows \(|\Psi(x,t)|^2\) for \(\gamma=10^{-2}\) where the values range from 0 (dark) to
        $\approx 3.62$ (light).  We use \(g = -0.1 J\) and the initial solitons have \(\psi(0)=1\) (\(N \approx 9\)), and \(|v(0)|
        \approx 0.5\) with opposite signs and are placed \(5w(0)\) apart on a space of size \(20w(0)\).  The figure shows that the
        interaction of the two solitons can enhance the energy dissipation.  }
    \label{fig:two}
\end{figure}

We display this in Fig.~\ref{fig:two} where we see that initially, i.e., before the interference begins (see inset), the
prediction agrees with the expectation of two separate solitons, i.e., the energy of the two solitons is twice the energy of the
single soliton.  During the evolution, when the solitons collide, the (unperturbed) energy of the two
solitons can even increase temporarily.
We ascribe this behaviour to interaction terms that go beyond the unperturbed single soliton [see Eq.~\eqref{eq:E_NSE}].
However, it is apparent that during the soliton collision energy is being dissipated beyond what would be
expected for separated
solitons.
We attribute this to the increase of the variation of the dissipation term  \(\propto \mathrm{Im} \left( \Psi^{*}
\frac{\partial \Psi}{\partial x^2} \right) \Psi\) as the solitons interfere with each other.

As a consequence, in a system of many solitons, energy is predominantly dissipated whenever two solitons collide. Beyond that,
there is the usual velocity-damping for each individual soliton, already discussed above.

\section{Conclusion}
\label{sec:conclusion}

In this article we have analyzed the behaviour of a bosonic many-body system subject to particle-conserving dissipation, treated
in the classical limit.
Particle-conserving dissipation is important in modern bosonic quantum simulators for platforms such as photonic systems that can be employed to simulate the interesting physical scenario of particle-conserving equilibration dynamics when subject to an engineered reservoir of this manner.
The results for the classical limit obtained here will provide a guide to understanding such experiments in the limit of large amplitudes. We have
derived a general equation of motion describing the effective dissipation. We have then applied it to the special case of a
Bose-Hubbard chain, where we have obtained a particular new version of a dissipative nonlinear Schrödinger equation.

We used numerics and a collective coordinate approach to study the behaviour of solitons in such a scenario and showed how
reservoir engineering can relax solitons to a stable shape as well as accelerate and de-accelarate moving solitons. Lastly we
briefly investigated the interaction of two solitons, where we found that extra energy is dissipated during the collision of
solitons.

Further works could build upon these insights to study the breakup of wave packets into soliton trains and the interactions of
multiple solitons as well as their long-term behaviour, all in the presence of particle-conserving dissipation. The behaviour of a
2D bosonic field coupled to this kind of dissipation remains completely unexplored territory as well.


\begin{acknowledgments}
  Benedikt Tissot acknowledges funding from the European Union’s Horizon 2020 research and innovation programme under Grant Agreement No.~862721 (QuanTELCO),
  as well as from the German Federal Ministry of Education and Research (BMBF) under the Grant Agreement No.~13N16212 (SPINNING).
\end{acknowledgments}


\begin{appendix}

\section{Classical dynamics using the Wirtinger derivative}
\label{app:wirtinger}

Considering a normal-ordered many-body Hamiltonian $\hat{H}_\uS$ of bosonic modes $\hat{b}_l$ ($l=1,\dots,L$), the Heisenberg
equations of motions can be expressed by \begin{align} i \dot{\hat{b}}_l & = \frac{1}{\hbar} \comm{\hat{b}_l}{\hat{H}_\uS} =
\frac{1}{\hbar} \sum_{n,m} \hat{A}_{nm} \comm{\hat{b}_l}{(\hat{b}_l^{\dag})^n} \hat{b}_l^m \notag \\ & = \frac{1}{\hbar}
\sum_{n,m} \hat{A}_{nm} n (\hat{b}_l^{\dag})^{n-1} \hat{b}_l^m, \end{align} where $\hat{A}_{nm}$ is made up by normal ordered
operators of all modes but $\hat{b}_l$.  If we now go to the classical limit $\hat{b}_l \to b_l$ we find the dynamics of the
classical amplitudes $b_l$ \begin{align} i \dot{b}_l & = \frac{1}{\hbar} \sum_{n,m} A_{nm} n (\hat{b}_l^{*})^{n-1} \hat{b}_l^m
\notag \\ & = \frac{1}{\hbar} \sum_{n,m} A_{nm} \frac{\partial (\hat{b}_l^{*})^{n}}{\partial b_l^{*}} \hat{b}_l^m =
\frac{1}{\hbar} \frac{\partial H_\uS}{\partial b_l^{*}}, \end{align} where $O = A_{nm}, H_\uS, $ corresponds to the classical version of $\hat{O}$
with the replacement $\hat{b}_n^{(\dag)} \to b_n^{(*)}$ for all modes $n$ and using the ``Wirtinger derivative''
$\partial/\partial b_l^*$ \cite{wirtinger1927formalen}.

\section{Integrating out the cavity}
\label{app:cavity}
In this appendix we present an alternative derivation of the effective dissipative dynamics specialized to the anharmonic chain.
The full Hamiltonian (in the frame rotating  with the bare frequency of the anharmonic sites and omitting the modes of the
environment only interacting with the bath modes $\hat{a}_i$) is given by:
\begin{align}
        \hat{H}_{\rm sys + bath} = \hbar \sum_{i=1}^L \Big[
        & - J \left( \hat{b}_{i+1}^{\dag} \hat{b}_i + \hat{b}_i^{\dag} \hat{b}_{i+1} \right) \notag \\
    \label{eq:H}
        & + \frac{\alpha}{2} \hat{b}_i^{\dag} \hat{b}_i (\hat{b}_i^{\dag} \hat{b}_i - 1)  \\
        & + \chi \hat{a}_i^{\dag} \hat{a}_i \hat{b}_i^{\dag} \hat{b}_i
        - \omega_c \left( \hat{a}_i^{\dag} \hat{a}_i + \frac{1}{2} \right)\Big] \notag
\end{align}
with the anihilation operators of the sites \(\hat{b}_i\) as well as the cavities \(\hat{a}_i\), the anharmonicity of the sites
\(\alpha\), the hopping rate \(J\), the cavity frequency \(\omega_c\), and the interaction strength \(\chi\).

Following the standard derivation for driven dissipative cavities (see \cite{clerk-2010-introd_to_quant_noise_measur_amplif} for a
review) and considering the regime where the quantum fluctuations are negligible compared to the classical amplitudes we find the
equations of motion of the classical amplitudes $A_n$ and $B_n$:
\begin{align}
    \label{eq:dA}
    \dot{A}_n = & i \Delta A_n + \eta - \frac{\kappa}{2} A_n  - i \chi |B_n|^2 A_n, \\
    \label{eq:dB}
    \dot{B}_n = & - i \chi |A_n|^2 B_n - i \alpha |B_n|^2 B_n + i J \left( B_{n-1} + B_{n+1} \right) ,
\end{align}
with the detuning \(\Delta = \omega_L - \omega_c\) between the cavity and the drive.  The terms proportional to the drive rate
$\eta$ and the decay rate $\kappa$ arise due to integrating out the environment following the standard procedure for a driven
dissipative cavity (with an interaction between the environmental modes $\hat{e}(\omega)$ with frequency $\omega$ and the cavity
 of the form $\hat{e}(\omega) \hat{a}^{\dag} + \mathrm{h.c.}$.  We note that in our notation all constants apart from
 \(\Delta,\alpha\) are positive.

First we note that only the hopping term leads to a change in the site occupation
\begin{equation}
  \frac{d \abs{B_n}^2}{dt} = i J \left( B_{n-1} + B_{n+1} \right) B_n^{*} + \mathrm{H.c.}
\end{equation}
which fulfills $| (d/dt) \abs{B_n}(t) |^2 \le J | B_{\mathrm{max}}(t) |^2$ with the maximal amplitude of the chain
$B_{\mathrm{max}}(t) = \max_n |B_n(t)|$.  For a stable soliton we can relate $B_{\mathrm{max}}(t) = \psi(t) =
\sqrt{N/2w_{\mathrm{SS}}}$, i.e. the square root of a density given by the number of excitations in the soliton devided by the
width of the soliton.
\begin{widetext}
Taking the ansatz
\begin{equation}
  A_n(t) = \frac{\eta}{- i \Delta + \kappa/2} + \exp \left[ (i \Delta - \kappa/2) t - i \chi \int_0^t \abs{B_n(t')}^2 dt' \right] {\delta A_{n}}(t)
  \label{app:eq:Aans}
\end{equation}
leads to
\begin{equation}
  \dot{\delta A_n} =  i \chi \frac{\eta}{i\Delta - \kappa/2} \abs{B_n(t)}^2 \exp \left[ (\kappa/2 - i \Delta) t + i \chi \int_0^t \abs{B_n(t')}^2 dt' \right] .
\end{equation}
Using \(\frac{d}{dt} \exp \left[ i \chi \int_0^t \abs{B_n(t')}^2 dt' \right] = i \chi \abs{B_n(t)}^2 \exp \left[ + i \chi \int_0^t
\abs{B_n(t')}^2 dt' \right]\) we find
\begin{equation}
  \dot{\delta A_n} = \frac{\eta}{i\Delta - \kappa/2} \exp \left[  (\kappa/2 - i \Delta) t \right] \left\{ \frac{d}{dt} \exp \left[ i \chi \int_0^t \abs{B_n(t')}^2 dt' \right] \right\} ,
\end{equation}
such that we can integrate both sides of the equation to find
\begin{equation}
    \begin{aligned}
        \delta A_n(t) = & \frac{\eta}{i\Delta - \kappa/2} \left\{ \exp \left[  (\kappa/2 - i \Delta) t + i \chi \int_0^t \abs{B_n(t')}^2 dt' \right] - 1 \right\}\\
        & + \eta \int_0^t \exp \left[  (\kappa/2 - i \Delta) t'' \right] \exp \left[ i \chi \int_0^{t''} \abs{B_n(t')}^2 dt' \right] dt'' .
    \end{aligned}
    \label{app:eq:varA}
\end{equation}
Where we approximate the integral \(\int_0^t e^{g(\tau)} d\tau\) with \(g(\tau) = ({\kappa}/{2} - i \Delta) \tau + i \chi
\int_0^{\tau} \abs{B_n(t')}^2 dt'\), therefore we first consider the magnitude of the derivatives \(|\dot{g}(t)| \ge
\abs{\kappa/2}\) and  \(|\ddot{g}(t)| < \chi J |B_{\mathrm{max}}|^2\) because \(\ddot{g}(t) = i \chi \frac{d \abs{B_n}^2}{d t}\) .
We use \(\frac{d}{dt} \frac{e^{g(t)}}{\dot{g}(t)} = e^{g(t)} - \frac{\ddot{g}(t)}{\dot{g}^2(t)} e^{g(t)}\) and  integrate by parts
\begin{equation}
  \int_0^t e^{g(\tau)} d\tau = \left[ \frac{e^{g(\tau)}}{\dot{g}(\tau)} \right]_0^t + \int_0^t \frac{\ddot{g}(\tau)}{\dot{g}^2(\tau)} e^{g(\tau)} d\tau ,
\end{equation}
where we can again  integrate by parts the second term
\begin{equation}
  \int_0^t \frac{\ddot{g}(\tau)}{\dot{g}^2(\tau)} e^{g(\tau)} d\tau = \left[ \frac{\ddot{g}(\tau)}{\dot{g}^3(\tau)} e^{g(t)}
  \right]_0^t - \int_0^t \left( \frac{\dddot{g}(\tau)}{\dot{g}^3(\tau)} - 2 \frac{\ddot{g}^2(\tau)}{\dot{g}^4(\tau)} \right)
  e^{g(\tau)} d\tau ,
\end{equation}
combined leading us to
\begin{equation}
  \int_{0}^t e^{g(\tau)} d\tau =  \left[ \frac{e^{g(\tau)}}{\dot{g}(\tau)} + \frac{\ddot{g}(\tau)}{\dot{g}^3(\tau)} e^{g(\tau)} \right]_0^t + \mathcal{O} \left[ \left( \frac{\dddot{g}(t)}{\dot{g}^5(t)} + \frac{\ddot{g}(t)}{\dot{g}^4(t)} \right) e^{g(t)} \right] .
\end{equation}
To get the correct order of these terms we consider
\begin{equation}
  \frac{1}{\dot{g}(t)} = \frac{1}{\kappa/2 - i\Delta + i \chi \abs{B_n}^2(t)} = \frac{1}{\kappa/2 - i\Delta} - \frac{i \chi \abs{B_n}^2(t)}{(\kappa/2 - i\Delta)^2} + \mathcal{O}\left[\frac{\chi^2 \abs{B_n}^4(t)}{(\kappa/2 - i\Delta)^3}\right] ,
\end{equation}
and
\begin{equation}
  \frac{\ddot{g}(t)}{\dot{g}^3(t)} = i \chi \frac{d \abs{B_n}^2}{d t}
 \left\{ \frac{1}{(\kappa/2 - i\Delta)^3} + \mathcal{O}\left[\frac{\chi \abs{B_n}^2(t)}{(\kappa/2 - i\Delta)^4}\right] \right\}.
\end{equation}
Therefore, we can approximate the second integral in \eqref{app:eq:varA} if \(\chi |B_{\mathrm{max}}|^2 \ll \abs{i \Delta +
\kappa/2}\) and \(\chi J |B_{\mathrm{max}}|^2 \ll \abs{i \Delta + \kappa/2}^2\) (i.e. in the weak coupling limit)
\begin{equation}
    \begin{aligned}
        \exp \left[ (i \Delta - \kappa/2) t - i \chi \int_0^t \abs{B_n(t')}^2 dt' \right] \delta A_n(t)
        = & \frac{\eta}{i\Delta - \kappa/2} \left\{ 1 - \exp \left[ (i \Delta - \kappa/2) t - i \chi \int_0^t \abs{B_n(t')}^2 dt' \right] \right\} \\
        & + \eta \left[ \frac{1}{\kappa/2 - i\Delta} - \frac{i \chi \abs{B_n}^2(t)}{(\kappa/2 - i\Delta)^2} + i \chi \frac{d \abs{B_n}^2}{d t} \frac{1}{(\kappa/2 - i\Delta)^3} \right] .
\end{aligned}
\end{equation}
Inserting this into \eqref{app:eq:Aans} and neglecting exponentially decaying terms we find
\begin{equation}
  A_n(t) \simeq
  \frac{\eta}{\kappa/2 - i\Delta} \left\{1  - \frac{i \chi}{\kappa/2 - i\Delta} \left[ \abs{B_n}^2  - \frac{\frac{d \abs{B_n}^2}{d t}}{\kappa/2 - i\Delta} \right] \right\} .
\end{equation}
Which we can square and approximate to the first order within the weak coupling approximation
\begin{equation}
  |A_n(t)|^2 \simeq
  \frac{\eta^2}{\kappa^2/4 + \Delta^2} \left\{1 + 2 \chi \left[ \frac{\Delta}{\kappa^2/4 + \Delta^2} \abs{B_n}^2 - \frac{\Delta \kappa}{(\kappa^2/4 + \Delta^2)^2} \frac{d \abs{B_n}^2}{d t} \right] \right\} .
  \label{app:eq:A2}
\end{equation}
Inserting this result into Eq.~\eqref{eq:dB} leads us to
\begin{equation}
  \dot{B}_n =  - i \chi \frac{\eta^2}{\kappa^2/4 + \Delta^2} \left\{1 + 2 \chi \left[ \frac{\Delta}{\kappa^2/4 + \Delta^2} \abs{B_n}^2 - \frac{\Delta \kappa}{(\kappa^2/4 + \Delta^2)^2} \frac{d \abs{B_n}^2}{d t} \right] \right\} B_n - i \alpha |B_n|^2 B_n + i J \left( B_{n-1} + B_{n+1} \right)
\end{equation}
Using the approximation in Eq.~\eqref{app:eq:A2} and changing into a rotating frame with \(b_n(t) = e^{i \chi \eta^2/(\Delta^2 + \kappa^2/4) t} e^{- i 2 J t} B_n(t)\),
we calculate
\begin{equation}
    \begin{aligned}
        \dot{b}_n
        = & \left( i \chi \frac{\eta^2}{\kappa^2/4 + \Delta^2} - 2 i J \right) b_n + e^{i \chi \eta^2/(\Delta^2 + \kappa^2/4) t} e^{i \pi n} e^{- i 2 J t} \dot{B_n} \\
        = & - i \gamma / 2 \frac{d \abs{b_n}^2}{d t} b_n - i g |b_n|^2 b_n + i J \left( b_{n-1} - 2 b_n + b_{n+1} \right)
    \end{aligned}
\end{equation}
with the effective self-interaction \(g = \alpha + \frac{2 \chi^2 \eta^2 \Delta}{(\kappa^2/4 + \Delta^2)^2}\) and the decay rate
\(\gamma J = - \frac{4 \chi^2 \eta^2 \Delta \kappa}{(\kappa^2/4 + \Delta^2)^3} J\) .  Finally we use the equation itself to
calculate
\begin{equation}
    \begin{aligned}
        \frac{d |b_n|^2}{dt}
        = & b_n^{*} \dot{b_n} + b_n \dot{b_n^{*}} = b_n^{*} \left[ - i \gamma \frac{d \abs{b_n}^2}{d t} b_n - i g |b_n|^2 b_n + i J \left( b_{n-1} - 2 b_n + b_{n+1} \right) \right] \\
        & + b_n \left[ i \gamma \frac{d \abs{b_n}^2}{d t} b_n^{*} + i g |b_n|^2 b_n^{*} - i J \left( b_{n-1}^{*} - 2 b_n^{*} + b_{n+1}^{*} \right) \right]
        = - 2 J \mathrm{Im} \left[ b_n^{*} \left( b_{n-1} - 2 b_n + b_{n+1} \right) \right],
    \end{aligned}
\end{equation}
which results in Eq.~\eqref{eq:dbEFF} of the main text,~i.e.
\begin{equation}
    i \dot{b}_n =  g |b_n|^2 b_n - J \left( b_{n-1} - 2 b_n + b_{n+1} \right) - J \gamma \mathrm{Im} \left[ b_n^{*} \left( b_{n-1} - 2 b_n + b_{n+1} \right) \right] b_n .
    \label{eq:dbEFFapp}
\end{equation}

\end{widetext}

\section{Particle Conserving NLSE}
\label{app:cons}

In this appendix we provide the short proof that
the \emph{particle conserving dissipative nonlinear Schrödinger Eq.~}\eqref{eq:nls} conserves the particle number.
We can show this by evaluating
\begin{equation}
    \begin{aligned}
        \frac{d}{dt} \int dx \abs{\Psi}^2(x,t)
        = & \int dx \frac{d}{dt} \abs{\Psi}^2(x,t) \\
        = & - J \int dx \mathrm{Im} \left( \Psi^{*} \frac{\partial^2 \Psi}{\partial x^2} \right) \\
        = & - J \mathrm{Im} \left[ \Psi^{*} \frac{\partial \Psi}{\partial x} \right]_0^L
        = 0,
    \end{aligned}
\end{equation}
where we used that we can add a real number inside the imaginary part to solve the integral, and that \(\Psi\) needs to either
vanish at the boundaries (for open boundary conditions) or \(\Psi(0) = \Psi(L)\) for periodic boundary conditions.  Therefore we
demonstrated that the particle number \(N = \int dx \abs{\Psi}^2(x,t)\) is constant.

\section{Variational Ansatz}
\label{app:var}

In this appendix we provide the necessary details to understand the derivation of the equations of motion for the collective
coordinates.  Without any coupling to the environment (\(\gamma=\chi=0\)) the \emph{particle conserving dissipative nonlinear
Schrödinger Eq.~}\eqref{eq:nls} becomes the standard unperturbed nonlinear Schrödinger equation which supports soliton solutions.
With this motivation we study the evolution of a generalized soltion ansatz
\begin{widetext}
\begin{equation}
    \bar{\Psi}(t) = \psi{(t)} \exp {\left\{ i \left[x - {x_{0}}{(t)}\right] v{(t)} + i \left[x - {x_{0}}{(t)}\right]^{2} d{(t)} + i \varphi{(t)} \right\}} \operatorname{sech}{\left[\frac{x - {x_{0}}{(t)}}{w(t)} \right]}
    \label{eq:ansAPP}
\end{equation}
\end{widetext}
within a variational approximation where the time-dependent parameters are called collective coordinates
\cite{scott-2006-nonlinear,dawes-2013-variat_approx_use_collec_coord}.  This approach was also employed to study soliton solutions
in different generalized non-linear Schrödinger equations in recent works
\cite{sahoo-2017-pertur_dissip_solit,rossi-2020-non_conser_variat_approx_nonlin_schroed_equat}.

Within this approach we first calculate the \emph{conservative} Lagrangian for the unperturbed nonlinear Schrödinger equation
\begin{equation}
    \bar{L} = \int_{-\infty}^{\infty} \left[ J {| \frac{\partial \bar{\Psi}}{\partial x} |}^2 + \frac{g}{2} {|\bar{\Psi}|}^4 +
    \frac{i}{2} \left( \bar{\Psi}^{*} \frac{\partial \bar{\Psi}}{\partial t} - \bar{\Psi} \frac{\partial \bar{\Psi}^{*}}{\partial t}
    \right) \right] dx ,
    \label{eq:Lconservative}
\end{equation}
and then we find the dynamics of the parameters of \(\bar{\Psi}\), i.e. \(p = \psi,x_0,d,v,\varphi,w\), using the perturbed
variation approach.  The resulting equation of motion are
\begin{equation}
    \frac{d}{dt} \frac{\partial \bar{L}}{\partial \dot{p}} - \frac{\partial \bar{L}}{\partial p} = 2 \Re \int_{- \infty}^{\infty} \bar{\mathcal{P}} \frac{\partial \bar{\Psi}^{*}}{\partial p} dx ,
    \label{eq:LagrangianEOM}
\end{equation}
with \(\bar{\mathcal{P}} =  - J \gamma \mathrm{Im} \left( \bar{\Psi}^{*} \frac{\partial^2 \bar{\Psi}}{\partial x^2} \right)\) (the
non-conservative term of the \emph{particle conserving dissipative nonlinear Schrödinger Eq.}).  Evaluating the integrals of the right
hand side combined with some algebra leads to the equations of motion of the main text \eqref{eq:EOMs1}--\eqref{eq:EOMs6}.

\section{Simulation Details}

All numerics are calculated using \cite{rackauckas-2017-diffeq.jl}.  In Fig.~\ref{fig:cont} we use the \texttt{Tsit5} algorithm
and relative (absolut) tolerance \(10^{-8}\) (\(10^{-8}\)) for the PCDNSE to achieve fast convergence for the large lattice (4001
points) simulating the continuous space.  The Langevin equations of motion have a much smaller lattice but are more prone to
numerical issues, therefore we use a \texttt{Vern9} algorithm with a relative tolerance of \(10^{-12}\).  In
Fig.~\ref{fig:velocity} (apart from one data point) use the same simulation setup used as Fig.~\ref{fig:cont}.  In
Fig.~\ref{fig:shape} we use the same simulation setup for the PCDNSE as Fig.~\ref{fig:cont}.  For the collective coordinates we
use a \texttt{Tsit5} with relative (absolute) tolerance \(10^{-10}\) (\(10^{-8}\)).  In Fig.~\ref{fig:two} we use the simulation
setup of Fig.~\ref{fig:cont} but a relative tolerance of \(10^{-10}\).

\end{appendix}

\bibliography{./refs.bib}
\end{document}